\documentclass[final,5p,times,twocolumn]{elsarticle}

\usepackage{amssymb}
\usepackage{amsthm}

\begin{document}

\begin{frontmatter}



\title{Dyakonov surface waves in lossy metamaterials}


\author[add1]{A.~J.~Sorni}
\author[add1]{M.~Naserpour}
\author[add1]{C.~J.~Zapata-Rodr\'{\i}guez\corref{cor1}}
\cortext[cor1]{Corresponding author\\ Email address: carlos.zapata@uv.es}
\author[add2]{J.~J.~Miret}

\address[add1]{Department of Optics and Optometry and Vision Science, University of Valencia, \\ C/ Dr. Moliner 50, Burjassot 46100, Spain}
\address[add2]{Department of Optics, Pharmacology and Anatomy, University of Alicante, P.O. Box 99, Alicante, Spain}

\begin{abstract}
We analyze the existence of localized waves in the vicinities of the interface between two dielectrics, provided one of them is uniaxial and lossy.
We found two families of surface waves, one of them approaching the well-known Dyakonov surface waves (DSWs).
In addition, a new family of wave fields exists which are tightly bound to the interface.
Although its appearance is clearly associated with the dissipative character of the anisotropic material, the characteristic propagation length of such surface waves might surpasses the working wavelength by nearly two orders of magnitude.
\end{abstract}


\end{frontmatter}





\section{Introduction}

Surface plasmons (SPs) appear in the disruption surface of isotropic media where the material permittivity changes of sign occurring with a metal in contact with a dielectric \cite{Maier07}.
The relevance of these surface plasmons falls not only upon its inherent subwavelength localization, but also they enable an amplification of evanescent signals traveling near the surfaces \cite{Fang03}.
These functionalities are being implemented within the lasts years for applications including optical sensing \cite{Lal07}, signal filtering \cite{Wang08}, subdiffraction focusing and subwavelength resolution imaging \cite{Fan06,Ceglia08,Miret10,Pastuszczak11,Zapata12a}.

Alternatively, it has been theoretically and experimentally demonstrated the existence of lossless surface waves at the interface of two different transparent dielectrics, provided one of them is anisotropic \cite{Dyakonov88,Takayama09}.
In opposition with SPs, this sort of surface waves have the peculiarity of possessing hybrid polarization \cite{Takayama08}.
The presence of hybrid surface waves with some parallel characteristics, additionally, may be found replacing the uniaxial medium by a biaxial crystal \cite{Walker98,Liscidini10}, an indefinite medium \cite{Yan07,Zapata13b}, and a structurally chiral material \cite{Gao09,Gao10}.
The use of structured materials with extreme anisotropy offered an alternative to increase the range of directions of DSWs substantially, as it is compared with the rather narrow range observed with natural birefringent materials \cite{Artigas05,Polo07}.
In particular, striking results are attained if the anisotropic structures include metallic nanoelements, as it occurs for example with a simple metal-dielectric (MD) multilayer, a case where the angular range may surpass half of a right angle \cite{Jacob08,Vukovic12}.
Caused by the specific damping capacity of metals, however, the propagation length of these DSWs is strongly limited by the penetration depth inside the lossy metamaterial \cite{Zapata13a}.

In this paper we perform a thorough analysis of DSWs taking place in lossy uniaxial metamaterials.
Special emphasis is put when the effective-medium approach (EMA) induces satisfactory results.
We examine lossy metamaterials that exhibit closed spatial-dispersion curves, in the same manner that occurs with natural birefringent crystals.
Contrarily, the introduction of losses leads to a transformation of the isofrequency curves, which deviates from spheres and ellipsoids, as commonly considered by ordinary and extraordinary waves, respectively.
As a consequence, two families of surface waves are found.
One family of surface waves are directly related with the well-known solutions derived by Dyakonov \cite{Dyakonov88}.
Importantly, we reveal the existence of a new family of surface waves, which is closely connected to the presence of losses in the uniaxial effective crystal.
The dominant diffusion dynamics of the latter surfaces waves is thoroughly examined.

\section{Theory and configuration}

\begin{figure}[htb]
 \centering
  \includegraphics[width=7.5cm]{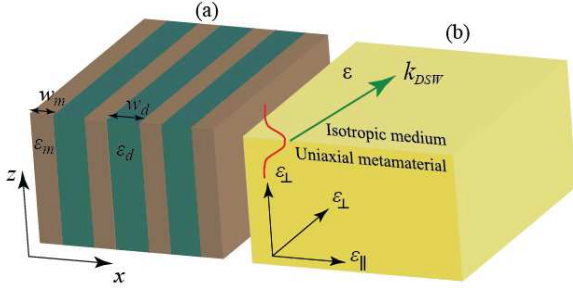}
 \caption{
  (a) Schematic arrangement under study, consisting of a multilayered metal-dielectric metamaterial ($z < 0$) and an isotropic lossless medium ($z > 0$) of dielectric constant $\epsilon$.
  (b) The effective anisotropy of the stratified metamaterial is represented by the on-axis permittivity, $\epsilon_{||}$, and the permittivity over its perpendicular direction, $\epsilon_\perp$.
  A Dyakonov surface wave propagates in the vicinities of the interface between the isotropic and the effective uniaxial medium with propagation constant $k_{DSW}$.
 }
 \label{fig01}
\end{figure}

We first analyze the dispersion properties of DSWs supported by a 1D array of semi-infinite layers made of a metal and a dielectric, displaced alternately as shown in Fig.~\ref{fig01}(a), where $w_m$ ($w_d$) is the width of the metallic (dielectric) layer within a given unit cell, and $\epsilon_m$ ($\epsilon_d$) represents the dielectric constant of this material. 
In the semi-space $z > 0$ we have a lossless dielectric environment of relative permittivity $\epsilon$.
An effective medium approach based on the long-wavelength approximation is used to calculate the permittivity of the anisotropic metamaterial along its optical axis $\epsilon_{||}$, that is the $x$-axis, and the permittivity in its normal direction, $\epsilon_\perp$.
An schematic of the modeled uniaxial metamaterial is shown in Fig.~\ref{fig01}(b).
Under these conditions, DSWs of propagation constant $k_{DSW}$ can be observed provided that $\epsilon_\perp < \epsilon < \epsilon_{||}$ \cite{Dyakonov88}.

\begin{figure}[htb]
 \centering
  \includegraphics[width=7.5cm]{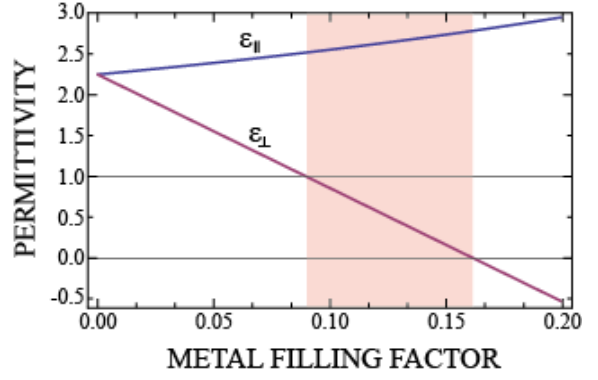}
 \caption{Effective permittivities $\epsilon_{||}$ and $\epsilon_\perp$ of a SiO$_2$-Ag multilayered metamaterial, at $\lambda_0 = 560\ \mathrm{nm}$, as a function of the metal filling factor $f$.
 In this simulation we disregard losses in the metamaterial, that is we used $\mathrm{Re}(\epsilon_{||})$ and $\mathrm{Re}(\epsilon_\perp)$.
 The shadowed region indicates the range of values of $f$ providing a necessary condition $0 < \epsilon_\perp < \epsilon < \epsilon_{||}$ for the existence of DSWs, where we considered $\epsilon = 1$ corresponding to air.}
 \label{fig02}
\end{figure}

Form birefringence in the anisotropic metamaterial is estimated in a simple way within the long-wavelength regime that enables an homogenization of the structured metamaterial \cite{Rytov56,Yariv77}.
The EMA demonstrates to be reliable for photonic structures including elements with sizes that are significantly smaller than the wavelength.
At infrared and visible wavelengths, the skin depth of noble metals is clearly subwavelength and, in this case, material homogenization requires that the metallic units had sizes of a few nanometers \cite{Elser07,Vukovic09,Chebykin11}.
In this case, the superlattice behaves as a uniaxial crystal whose optical axis is normal to the layers.
The model estimates the relative permittivities along the optical axis, 
\begin{equation}
 \epsilon_{||} = \frac{\epsilon_m \epsilon_d}{\left( 1 - f \right) \epsilon_m + f \epsilon_d} ,
\label{eq01}
\end{equation}
and transversally, 
\begin{equation}
 \epsilon_\perp = \left( 1 - f \right) \epsilon_d + f \epsilon_m ,
\label{eq02}
\end{equation}
where $f = {w_m}/\left({w_d + w_m}\right)$ is the metal filling factor.
If we neglect losses by setting $\mathrm{Im}(\epsilon_{||}) = 0$ and $\mathrm{Im}(\epsilon_\perp) = 0$, the effective birefringence of the MD superlattice is $\Delta n = \sqrt{\epsilon_{||}} - \sqrt{\epsilon_\perp}$.
A small filling factor of the metallic composite may lead to an enormous birefringence, as inferred from Fig.~\ref{fig02}.

The effective permittivities of a MD periodic multilayer made of silicon dioxide ($\epsilon_d = 2.25$) and silver ($\epsilon_m = -11.7 + i 0.83$ taken from \cite{Palik99}) at a wavelength $\lambda_0 = 560\ \mathrm{nm}$ are displayed in Fig.~\ref{fig02}, showing how they vary for a different metal filling factor $f$.
For the sake of clarity, we ignored the dissipative effects in the uniaxial metamaterial, and here we considered the real part of $\epsilon_{||}$ and $\epsilon_\perp$ obtained from Eqs.~(\ref{eq01}) and (\ref{eq02}).
In this case, these are real and positive permittivities provided that $f < 0.161$, in addition leading to positive birefringence.
Otherwise, the permittivity $\epsilon_\perp$ became negative for a higher metal filling factor.
In addition, we considered cases where $f > 0.0896$, that is uniaxial metamaterials whose perpendicular permittivity satisfies $\epsilon_\perp < \epsilon$ enabling the existence of DSWs; in this numerical simulation we considered $\epsilon = 1$ corresponding to air.

\begin{figure}[htb]
 \centering
  \includegraphics[width=5.5cm]{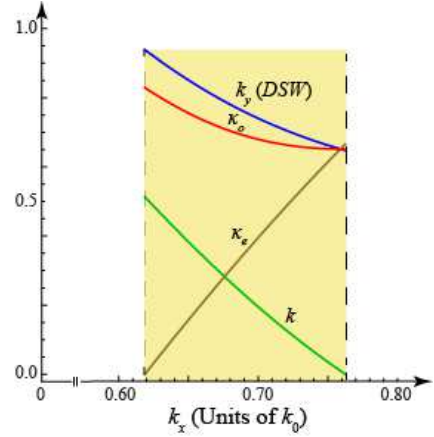}
 \caption{Isofrequency curve in the plane $k_x$-$k_y$ of the solutions of the Dyakonov equation (\ref{eq03}) calculated for $\epsilon_{||} = 2.625$, $\epsilon = 1$, and $\epsilon_\perp = 0.576$, at a wavelength $\lambda_0 = 650\ \mathrm{nm}$.
 We include the dependence of $\kappa$, $\kappa_o$, and $\kappa_e$ on the spatial frequency $k_x$. 
 The spatial spectrum that corresponds to DSWs is bounded by $k_x = 0.619 k_0$ ($\kappa_e = 0$) and $k_x =  0.763 k_0$ ($\kappa = 0$).}
 \label{fig03}
\end{figure}

In Fig.~\ref{fig03} we represent the solutions of the Dyakonov equation, namely
\begin{equation}
 \left( \kappa + \kappa_e \right) \left( \kappa + \kappa_o \right) \left( \epsilon \kappa_o + \epsilon_\perp \kappa_e \right)
  = \left( \epsilon_{||} - \epsilon \right) \left( \epsilon - \epsilon_\perp \right) k_0^2 \kappa_o ,
\label{eq03}
\end{equation}
providing the in-plane wave vector $(k_x,k_y)$ of the hybrid-polarized surface wave; note that the modulus of such a wave vector is the propagation constant $k_{DSW}$ of the DSW.
The electromagnetic fields are evanescent in the isotropic medium, proportional to $\exp \left( - \kappa |z| \right)$, where $k_0 = 2 \pi / \lambda_0$.
On the other side of the boundary, the ordinary and extraordinary waves in the effective uniaxial medium also decay exponentially with rates given by $ \kappa_o $ and $ \kappa_e $, respectively.
To illustrate the characteristic dispersion of DSWs, we solved Eq.~(\ref{eq03}) for the uniaxial SiO$_2$-Ag metamaterial described above, where the metal filling factor $f = 0.12$, which is covered by air; again we assumed that the wavelength $\lambda_0 = 650\ \mathrm{nm}$.
In this case, DSWs propagate within a wide angular region $\Delta \varphi = \varphi_\mathrm{max} - \varphi_\mathrm{min}$, where $\varphi$ is the polar angle of the surface wave vector $(k_x,k_y)$.
Specifically, we estimate $\Delta \varphi = 16.3^\circ$ around a mean angle $\bar \varphi = 48.5^\circ$ (DSWs lying in the first quadrant).
The dispersion curve for DSWs is drawn in Fig.~\ref{fig03}.
As shown in that figure, $\varphi_\mathrm{min}$ is attained under the condition $\kappa = 0$ (green solid line).
On the other side of the dispersion curve, $\varphi_\mathrm{max}$ is determined by $\kappa_e = 0$ (brown solid line) for which the extraordinary wave ceases to attenuate at $x \to + \infty$ \cite{Dyakonov88,Walker98}.

\section{Simulation results and discussion}

Metal losses were disregarded in the analysis given above.
In a realistic examination, however, the permittivity $\epsilon_m$ of the metal is not a real-valued constant, thus leading to a non-vanishing imaginary part of the permittivities $\epsilon_{||}$ and $\epsilon_\perp$ of the metallic compound.
The same occurs for the spatial frequencies $k_x$ and $k_y$ that are solutions of Eq.~(\ref{eq03}).
The following procedure can be used to find solutions of the Dyakonov equation: for a given real-valued spatial frequency $k_x$ along the optic-axis of the metamaterial, we calculated the complex-valued spatial frequencies $k_y$ that satisfies Eq.~(\ref{eq03}).
This method is valid provided that the excitation of the surface waves, given in terms of a plane-wave Fourier expansion, are produced at a plane parallel to $y = 0$, and that $ \mathrm{Im} (k_y) > 0$ \cite{Zapata13a}.
As a result of utilizing a lossy uniaxial material, surface waves cannot propagate indefinitely and DSWs decay with a propagation length given by \cite{Vukovic12,Warmbier12}
\begin{equation}
 L_{DSW} = \left[ 2 \mathrm{Im} (k_y) \right]^{-1} .
\label{eq04}
\end{equation}

\begin{figure}[htb]
 \centering
  \includegraphics[width=7.5cm]{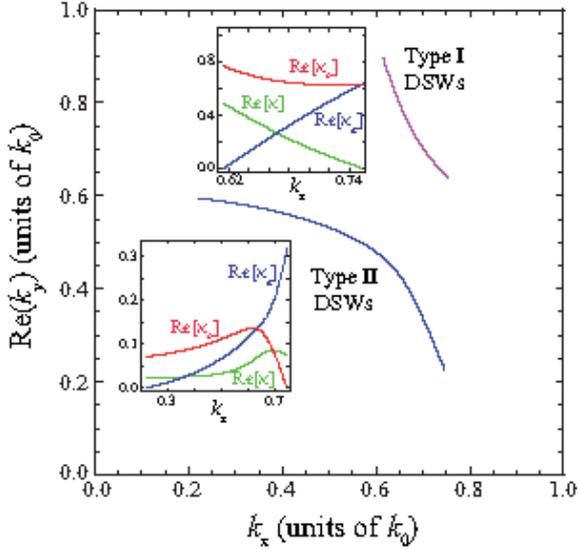}
 \caption{Dispersion curves for solutions of Dyakonov equation (\ref{eq03}) considering a lossy metamaterial compound of permittivities $\epsilon_{||} = 2.625 + i 0.005$ and $\epsilon_\perp = 0.576 + i 0.100$ at a wavelength $\lambda_0 = 560\ \mathrm{nm}$.
 We observe the existence of two families of hybrid-polarized surface waves, for which the real part of $\kappa$, $\kappa_o$, and $\kappa_e$ has a positive value, as shown in the insets.}
 \label{fig04}
\end{figure}

In Fig.~\ref{fig04} we represent the spatial frequency $\mathrm{Re} (k_y)$ of dissipative DSWs propagating at different on-axis spatial frequencies $k_x$ over a lossy SiO$_2$-Ag metamaterial, where the metal filling factor $f=0.12$.
Localization of the wave field in the vicinities of the plane $z = 0$ involves that the real part of $\kappa$, $\kappa_o$, and $\kappa_e$ has a positive value, as shown in the insets of Fig.~\ref{fig04}.
In principle, no restrictions must be imposed to their imaginary part \cite{Burke86}.
Physically-realizable solutions of Eq.~(\ref{eq03}) are given within two spectral domains.
The first domain has a spatial bandwidth that is comprised between $k_x = 0.615 k_0$ and $k_x = 0.753 k_0$, where $\mathrm{Re}(k_y)$ takes values from $0.639 k_0$ and $0.895 k_0$, which is in good correspondence to the DSWs shown in Fig.~\ref{fig03}.
Interestingly, a second domain for Dyakonov solutions emerges within a higher bandwidth (of the spatial frequency $k_x$) ranging from $0.220 k_0$ to $0.746 k_0$.
This new family of surface waves, here coined as type II DSWs, turns out to be the main result of our study.

\begin{figure}[htb]
 \centering
  \includegraphics[width=7.5cm]{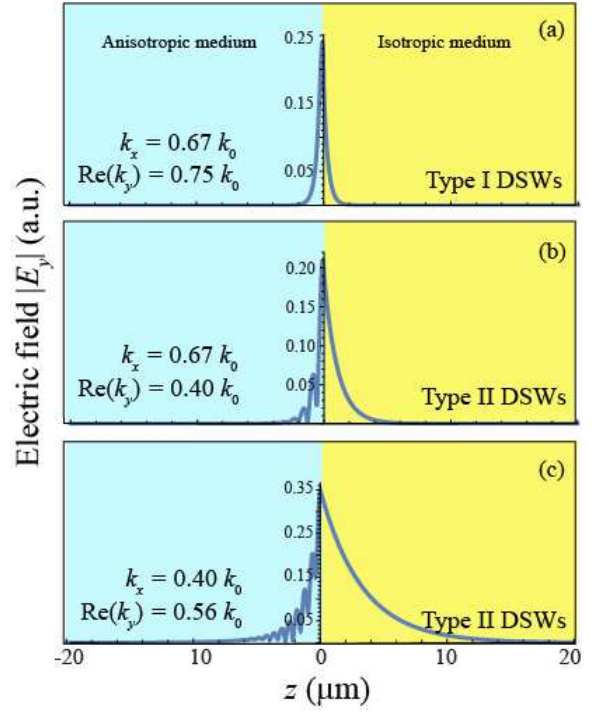}
 \caption{Electric field $|E_y|$, in arbitrary units, in the vicinities of the interface $z=0$ for Dyakonov-type surface waves propagating at a spatial frequency: (a) $k_x = 0.67 k_0$ associated with the first family of DSWs, (b) $k_x = 0.67 k_0$ and (c) $k_x = 0.40 k_0$ corresponding the the new family of surface waves.}
 \label{fig05}
\end{figure}

Figure \ref{fig05} depicts the modulus of the $y$-component of the electric field, $|E_y|$, in the vicinities of the interface $z=0$ for hybrid-polarized surface waves propagating at a spatial frequency: (a) $k_x = 0.67$ ($k_y = 0.752 + i 0.0847$ expressed in units of $k_0$) associated with (type I) DSWs, (b) $k_x = 0.67$ ($k_y = 0.397 + i 0.136$) and (c) $k_x = 0.40$ ($k_y = 0.562 + i 0.035$) corresponding the the new family of surface waves (type II DSWs) that we found in our study.
The $x$- and $z$-component of the electric field (not shown in the figure) are also localized around ($z = 0$).
The analytical expression of the electric field for type II DSWs is the same as the expression that describes type I DSWs, which can be found for instance in Refs.~\cite{Takayama08} and \cite{Zapata13b}.
We observe a strong field confinement near the isotropic-anisotropic interface in all cases, though type I DSWs present a superior performance.
This fact can also be inferred from the higher values of the evanescent decay rates calculated for type I and type II DSWs, which are shown in the insets of Fig.~\ref{fig04}.
In addition, a modulation on the tail of the evanescent field inside the lossy anisotropic medium is evident.
This is caused by the existence of a non-vanishing imaginary part of the decay rates $\kappa_o$ and $\kappa_e$; on the contrary such a modulation cannot be observed in the isotropic medium since field penetration is performed at a single rate $\kappa$.
Finally, the modulation of the field intensity is proportional to $\cos [\mathrm{Im} (\kappa_o - \kappa_e) x]$, which can be observed provided that its period is shorter than the penetration depth in the medium, as shown in Figs.~\ref{fig05}(b) and (c).

\begin{figure}[htb]
 \centering
  \includegraphics[width=7.5cm]{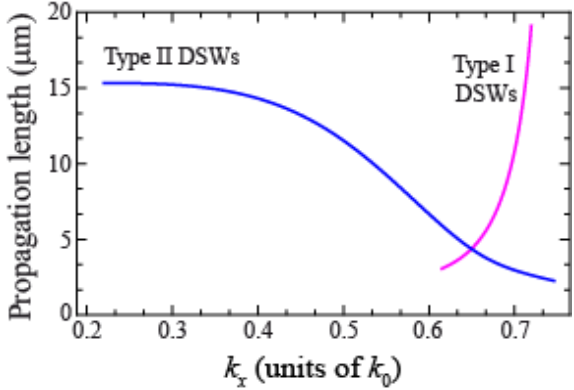}
 \caption{Propagation length of type I DSWs and type II DSWs.}
 \label{fig06}
\end{figure}

In Fig.~\ref{fig06} we show the propagation length, defined in Eq.~(\ref{eq04}), and evaluated for type I DSWs and type II DSWs. 
Contrarily to what one may know by intuition, type II DSWs can propagate for a longer distance than type I DSWs, occurring in the spectral interval comprised between $k_x = 0.62 k_0$ and $k_0 = 0.65 k_0$.
Moreover, the propagation length of type II DSWs can even surpasses $15\ \mu\mathrm{m}$ at low values of the spatial frequency $k_x$.
This is caused by the fact that the real part of $\kappa$ is very low within this spectral region, leading to low-decaying evanescent tails in the isotropic medium; see Fig.~\ref{fig05}(c) as an illustration of such an effect.
Although the peak is placed close to the surface $z = 0$, in this case the average location of the field intensity is severely shifted toward the lossless material.
Note that this also happens in long-range surface plasmons polaritons \cite{Berini09}.

\section{Conclusions}

We numerically investigated the dispersion properties of Dyakonov surface waves propagating on the interface between an isotropic loss-free medium and a lossy uniaxial crystal, the latter resulting by applying the long-wavelength approach to metal-dielectric multilayered metamaterials.
We conclude that the presence of absorption in the anisotropic medium limits the propagation length of these surface waves, but the spatial dispersion and field distribution near the material discontinuity varies slightly.
In addition, we found a new family of electromagnetic waves which are confined near the isotropic-anisotropic interface and propagate obliquely to the optic axis of the birefringent material.
These surface waves, here coined as type II Dyakonov surface waves, cannot be found in a configuration using lossless materials.
Unexpectedly, the propagation length of such surface wave fields might surpasses that related to type I DSWs.
Potential applications of type II DSWs include ultra-compact waveguiding and biosensing.

\section*{Acknowledgments}

This research was funded by the Spanish Ministry of Economy and Competitiveness under the project TEC2013-50416-EXP.








\end{document}